\documentclass[conference]{IEEEtran}
\makeatletter
\def\ps@headings{%
\def\@oddhead{\mbox{}\scriptsize\rightmark \hfil \thepage}%
\def\@evenhead{\scriptsize\thepage \hfil \leftmark\mbox{}}%
\def\@oddfoot{}%
\def\@evenfoot{}}
\makeatother
\usepackage{graphicx}
\usepackage{amssymb}
\usepackage{epstopdf}
\usepackage{amsmath}
\usepackage{enumerate}
\usepackage{cite}
\usepackage{verbatim}
\usepackage{mathcomp}
\usepackage{supertabular}
\usepackage{longtable}
\usepackage{stmaryrd}
\usepackage{url}
\newcommand{\C}{{\mathcal C}}

\newcommand{\G}{{\mathcal G}}

\newcommand{\X}{{\mathcal X}}

\newcommand{\bM}{{\mathbf M}}

\newcommand{\entropy}{{\sf H}}

\newcommand{\vc}{{\mathbf c}}
\newcommand{\vcone}{{\mathbf c}^{(1)}}
\newcommand{\vctwo}{{\mathbf c}^{(2)}}
\newcommand{\vci}{{\mathbf c}^{(i)}}
\newcommand{\vcj}{{\mathbf c}^{(j)}}
\newcommand{\vck}{{\mathbf c}^{(k)}}

\newcommand{\vs}{{\mathbf s}}

\newcommand{\vu}{{\mathbf u}}

\newcommand{\vv}{{\mathbf v}}
\newcommand{\vvone}{{\mathbf v}^{(1)}}
\newcommand{\vvtwo}{{\mathbf v}^{(2)}}

\newcommand{\vvj}{{\mathbf v}^{(j)}}

\newcommand{\vx}{{\mathbf x}}

\newcommand{\vy}{{\mathbf y}}
\newcommand{\vz}{{\mathbf z}}
\newcommand{\vO}{{\mathbf 0}}

\newcommand{\ve}{{\mathbf e}}

\newcommand{\vX}{\mathbf{X}}

\newcommand{\vG}{\mathbf{G}}

\newcommand{\vY}{\mathbf{Y}}

\newcommand{\bw}{{\mathbf{w}}}
\newcommand{\bg}{{\mathbf{g}}}

\newcommand{\rank}{{\rm rank}}
\newcommand{\supp}{{\rm supp}}

\newcommand{\bbN}{{\mathbb N}}

\newcommand{\al}{\alpha}

\newcommand{\be}{\beta}

\newcommand{\dist}{{\mathsf{d}}}
\newcommand{\weight}{{\mathsf{wt}}}
\newcommand{\spn}{{\mathsf{span}}}

\newcommand{\define}{\stackrel{\mbox{\tiny $\triangle$}}{=}}

\renewcommand{\ge}{\geqslant}
\renewcommand{\le}{\leqslant}

\newcommand{\mc}{\mathcal{C}}
\newcommand{\ff}{\mathbb{F}}
\newcommand{\fq}{\mathbb{F}_q}
\newcommand{\fqn}{\mathbb{F}_q^n}
\newcommand{\et}{{\emph{et. al.}}}

\newtheorem{definition}{Definition}[section]
\newtheorem{example}{Example}[section]
\newtheorem{theorem}{Theorem}[section]

\newtheorem{lemma}[theorem]{Lemma}

\newtheorem{corollary}[theorem]{Corollary}

\begin{document}

\title{Secure Index Coding with Side Information}

\author{
  \IEEEauthorblockN{Son Hoang Dau, Vitaly Skachek, and Yeow Meng Chee}
  \IEEEauthorblockA{Division of Mathematical Sciences,
    School of Physical and Mathematical Sciences\\
    Nanyang Technological University,
    21 Nanyang Link, Singapore 637371\\
    Emails: {\tt \{ DauS0002, Vitaly.Skachek, YMChee \} @ntu.edu.sg}}
}

\maketitle

\begin{abstract}
\boldmath 
Security aspects of the Index Coding with Side Information (ICSI) problem are investigated. 
Building on the results of Bar-Yossef \emph{et al.} (2006), the properties of linear coding 
schemes for the ICSI problem are further explored. 
The notion of weak security, considered by Bhattad and Narayanan (2005) in the context of network coding, is 
generalized to \emph{block security}. 
It is shown that the coding scheme for the ICSI problem based on a linear code $\mc$ of length $n$, minimum distance $d$ and dual distance $d^\perp$, is $(d-1-t)$-block secure (and hence also weakly secure) if the adversary knows in advance $t \leq d-2$ messages, and is completely insecure if the adversary knows in advance more than $n - d^\perp$ messages. 
\end{abstract}


\section{Introduction}
\label{sec:introduction}

\PARstart{T}he problem of Index Coding with Side Information (ICSI) was introduced by Birk and Kol~\cite{BirkKol98}, \cite{BirkKol2006}. It was motivated by applications such as audio and video-on-demand, and daily newspaper delivery. In these applications a server (sender) has to deliver some sets of data, audio or video files to the set of clients (receivers), 
different sets are requested by different receivers. Assume that before the transmission starts, 
the receivers have already (from previous transmissions) some files or movies in their possession.
Via a slow backward channel, the receivers can let the sender know which messages they already have in their possession, and which messages they request. By exploiting this information, the amount of the overall transmissions can be reduced. 
As it was observed in~\cite{BirkKol98}, this can be achieved by coding the messages at the server 
before broadcasting them out. 

Another possible application of the ICSI problem is in opportunistic wireless networks. 
These are the networks in which a wireless node can opportunistically listen to the wireless channel. As a result, the node may obtain packets that were not designated to it (see~\cite{Rouayheb2009, Katti2006, Katti2008}). This way, a node obtains some side information about the transmitted data. Exploiting this additional knowledge may help to increase the throughput of the system. 

Consider the toy example in Figure~1. It presents a scenario with one sender and four receivers. 
Each receiver requires a different information packet (or message). 
The na\"ive approach requires four separate transmissions, one transmission per an information
packet. However, by exploiting the knowledge about the subsets of messages that clients already have, and by using coding of the transmitted data, the server can satisfy all the demands by broadcasting just one coded packet. 

The ICSI problem has been a subject of several recent studies \cite{Yossef, LubetzkyStav, Wu, Rouayheb2007, Rouayheb2008, Rouayheb2009, ChaudhrySprintson, Alon}. This problem can be regarded as a special case of the well-known network coding (NC) problem~\cite{Ahlswede}. In particular, it was shown that every instance of the NC problem can be reduced to an instance of the ICSI problem~\cite{Rouayheb2008, Rouayheb2009}.

\begin{figure}
\begin{center}
\includegraphics[scale = 0.12]{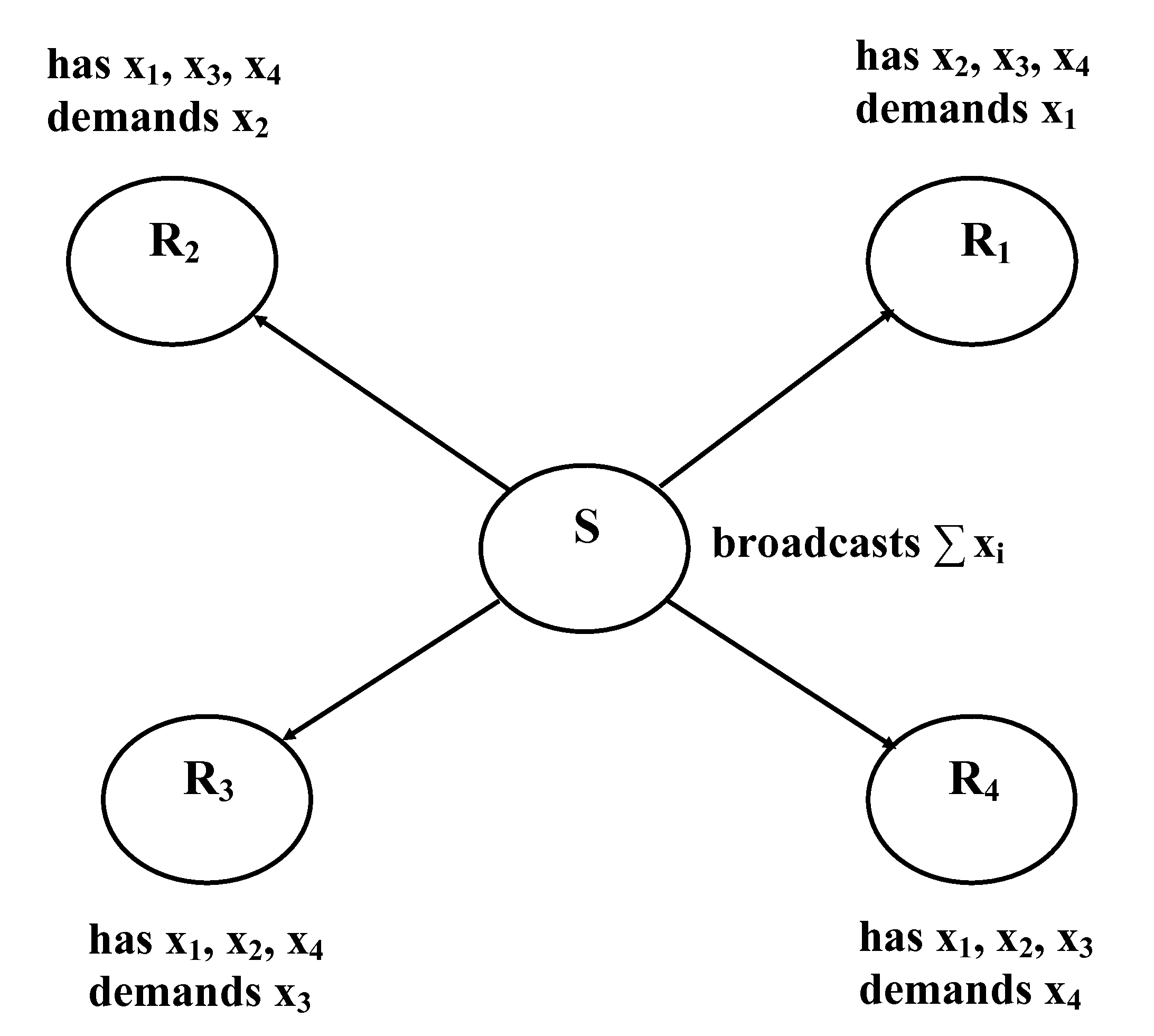}
\end{center}
\caption{An example of the ICSI problem}  
\end{figure}

Several previous works focused on the design of an efficient scheme for the ICSI problem. Bar-Yossef \emph{et al.} \cite{Yossef} proved that finding the best \emph{scalar linear binary solution} for the ICSI problem is equivalent to finding the so-called \emph{minrank} of a graph, which is known to be an NP-hard problem (see \cite{Yossef, Peeters96}). Here scalar linear solutions refer to linear schemes in which each message is a symbol in the field $\fq$. By contrast, in \emph{vector linear solutions} each message is a vector over $\fq$. Lubetzky and Stav \cite{LubetzkyStav} showed that there exist instances in which scalar linear solutions over nonbinary fields and linear solutions over mixed fields outperform the scalar linear binary solutions. The latter were also studied by Bar-Yossef \emph{et al.} \cite{Yossef}. 

El Rouayheb \et \cite{Rouayheb2008, Rouayheb2009} and Alon \emph{et al.} \cite{Alon} showed that for certain instances of the ICSI problem, vector linear solutions achieve strictly higher transmission rate than scalar linear solutions do. They also pointed out that there exist instances in which nonlinear codes outperform linear codes. Several heuristic solutions for the ICSI problem were proposed in \cite{Rouayheb2007, ChaudhrySprintson}. 
          
In this paper, we study the security aspect of a linear solution for the ICSI problem. 
We show that every linear scheme provides a certain level of security. More specifically, let $n$ and $k$ be the 
length and the dimension of the code $\mc$, associated with a particular ICSI instance. 
Let $d$ and $d^\perp$ be its minimum distance and dual distance, respectively. 
We say that a particular adversary is of strength $t$ if it has $t$ packets of information in its possession. 
Then, we show that a scheme employing the code $\mc$ is $(d-1-t)$-block secure against all adversaries 
of strength $t \leq d - 2$ and is completely insecure against any adversary of strength at least 
$n - d^\perp + 1$. If the code $\mc$ is MDS, then the two bounds coincide. 

The paper is organized as follows. Notations and definitions, which are used in the rest of the paper, are introduced in Section~\ref{sec:preliminaries}. The model and some basic results for the ICSI problem are presented in Section \ref{sec:icsi}. The security properties of linear ICSI schemes are analyzed in Section~\ref{sec:our-results}. 
The main results of this paper appear in that section. Finally, the paper is concluded in Section~\ref{sec:conclusion}.  

\section{Preliminaries}
\label{sec:preliminaries}

We use the notation $\fq$ for the finite field of $q$ elements, where $q$ is a power of prime, and $\fq^*$ for the set of all nonzero elements of $\fq$. We also use $[n]$ to denote the set of integers $\{1,2,\ldots,n\}$.
For the vectors $\vu = (u_1, u_2, \ldots, u_n) \in \fq^n$ and $\vv = (v_1, v_2, \ldots, v_n) \in \fq^n$, 
the (Hamming) distance between $\vu$ and $\vv$ is defined to be the number of coordinates where $\vu$ and $\vv$ differ, 
namely, 
\[
\dist(\vu,\vv) = |\{i \in [n] \; : \; u_i \ne v_i\}|  \; . 
\]
The \emph{support} of a vector $\vu \in \fqn$ is defined to be the set $\text{supp}(\vu) = \{i \in [n]: u_i \ne 0\}$.  
The (Hamming) weight of a vector $\vu$, denoted $\weight(\vu)$, is defined to be $|\text{supp}(\vu)|$, the number of nonzero coordinates of $\vu$. 

A $k$-dimensional subspace $\mc$ of $\fq^n$ is called a linear $[n,k,d]_q$ ($q$-ary) code if the minimum distance of $\mc$, 
\[
\dist(\mc) \define \min_{\vu \in \mc, \; \vv \in \mc, \; \vu \neq \vv} \dist(\vu,\vv) \; ,
\]
is equal to $d$. Sometimes we may use the notation $[n,k]_q$ for the sake of simplicity. The vectors in $\mc$ are called codewords. It is easy to see that the minimum weight of a nonzero codeword in a linear code $\mc$ is equal to its minimum distance $\dist(\mc)$. A \emph{generator matrix} $\vG$ of an $[n,k]_q$-code $\mc$ is a $k \times n$ matrix whose rows are linearly independent codewords of $\mc$. Then $\mc = \{\vy \vG: \vy \in \fq^k\}$. 

The dot product of the two vectors $\vu, \vv \in \fq^n$ is defined to be $\vu \cdot \vv = \sum_{i=1}^n u_i v_i \in \fq$. Thus, $\vu \cdot \vv = \vu \vv^T$, the normal matrix product of $\vu$ and $\vv^T$, where $\vv^T$ denotes the transpose of $\vv$. The \emph{dual code} or \emph{dual space} of $\mc$ is defined as $\mc^\bot = \{\vu \in \fq^n \; : \; \vu \cdot \vc = \vO \text{ for all } \vc \in \mc\}$. The minimum distance of $\mc^\bot$, $\dist(\mc^\bot)$, is called the dual distance of $\mc$.     

The following upper bound on the minimum distance of a $q$-ary linear code is well-known (see~\cite{MW_and_S} Chapter 1).

\medskip
\begin{theorem}[Singleton bound]
\label{singleton}
For an $[n,k,d]_q$-code, we have $d \leq n - k + 1$. 
\end{theorem}
\medskip 

Codes attaining this bound are called \emph{maximum distan\-ce se\-pa\-rable} (MDS) codes. For a subset of vectors 
$$
\{ \vcone, \vctwo,\ldots,\vck \} \subseteq \fq^n \; , 
$$ define its linear span:  
\begin{multline*}
\spn\left( \{ \vcone,\vctwo, \ldots, \vck \} \right) \define \\ \left\{ \sum_{i = 1}^k \al_i \vci \; : 
\; \al_i \in \fq, \; i \in [k] \right\} \; .
\end{multline*}
We use $\ve_i = (\underbrace{0,\ldots,0}_{i-1},1,\underbrace{0,\ldots,0}_{n-i}) \in \fqn$ to denote the unit vector, which has a one at the $i$th position, and zeros elsewhere. We also use $\mathbf{I}_n$, 
$n \in \bbN$, to denote the $n \times n$ identity matrix.   

We recall the following well-known result in coding theory.

\medskip
\begin{theorem}[\cite{Hedayat}, p. 66]
\label{thmOA}
Let $\mc$ be an $[n,k,d]_q$-code with dual distance $d^\perp$ and $\bM$ denote the $q^k \times n$ matrix whose $q^k$ rows are codewords of $\mc$. If $r \leq d^\perp-1$ then each $r$-tuple from $\fq$ appears in an arbitrary set of $r$ columns of $\bM$ exactly $q^{k-r}$ times. 
\end{theorem}
\medskip

For a random vector $\vY = (Y_1,Y_2,\ldots,Y_n)$ and a subset $B = \{i_1,i_2,\ldots,i_b\}$ of $[n]$, where $i_1 < i_2 < \ldots <i_b$, let $\vY_B$ denote the vector $(Y_{i_1},Y_{i_2},\ldots,Y_{i_b})$. For a $k \times n$ matrix $\bM$, let $\bM_j$ denote the $j$th column of $\bM$, and $\bM[i]$ its $i$th row. For a set $E \subseteq [n]$, let $\bM_E$ denote the $k \times |E|$ matrix obtained from $\bM$ by deleting all the columns of $\bM$ which are not indexed by the elements of $E$. 

Let $X$ and $Y$ be discrete random variables taking values in the sets $\Sigma_X$ and $\Sigma_Y$, respectively. Let $\text{Pr}(X=x)$ denote the probability that $X$ takes a particular value $x \in \Sigma_X$. 
The (binary) \emph{entropy} of $X$ is defined as
\[
\entropy_2(X) = -\sum_{x \in \Sigma_X}\text{Pr}(X = x) \cdot \log_2\text{Pr}(X = x) \; . 
\]
The \emph{conditional entropy} of $X$ given $Y$ is defined as 
\begin{multline*}
\hspace{-2ex} \entropy_2(X|Y) = \\
-\sum_{x \in \Sigma_X, y \in \Sigma_Y} \text{Pr}(X = x,Y = y) 
\cdot \; \log_2 \text{Pr}(X = x|Y = y) \; . 
\end{multline*}
This definition can be naturally extended to 
$$
\entropy_2(X|Y_1, Y_2, \ldots, Y_n) \; , 
$$ 
for $n$ discrete 
random variables $Y_i$, $i \in [n]$.

If the probability distribution of $X$ is unchanged given the knowledge of $Y$, i.e., $\text{Pr}(X = x|Y  = y) = 
\text{Pr}(X = x)$ for all $x \in \Sigma_X$, $y \in \Sigma_Y$, 
then $\entropy_2(X|Y) = \entropy_2(X)$. Indeed, $\entropy_2(X|Y)$ equals 
\begin{eqnarray*}
&&\hspace{-9ex} - \; \sum_{x\in \Sigma_X} \left(\sum_{y \in \Sigma_Y}\text{Pr}(X = x, Y = y)\right) \cdot \log_2 \text{Pr}(X = x)\\
&=&-\sum_{x\in \Sigma_X} \text{Pr}(X = x) \cdot \log_2 \text{Pr}(X = x)\\
&=&\entropy_2 (X) \; . 
\end{eqnarray*} 
 
\section{Index Coding and Some Basic Results}
\label{sec:icsi}

\subsection{Linear Coding Model}
\label{sec:linear-icsi} 

Index coding problem considers the following communications scenario. There is a unique sender (or source) $S$, who has a vector of messages $\vx = (x_1, x_2, \ldots, x_n) \in \fqn$ in his possession, which is a realized value of a random vector $\vX = (X_1,X_2,\ldots,X_n)$. $X_1,X_2,\ldots,X_n$ hereafter are assumed to be independent uniformly distributed 
random variables over $\fq$. 
There are also $m$ receivers $R_1,R_2,\ldots,R_m$. For each $j \in [m]$, $R_j$ has some side information, 
i.e. $R_j$ owns a subset of messages $\{ x_i \}_{i \in \X_j}$, $\X_j \subseteq [n]$. In addition, each $R_j$, $j \in [m]$, 
is interested in receiving the message $x_{f(j)}$, for some \emph{demand function} $f \, : \, [m] \rightarrow [n]$. Hereafter, we assume that every receiver requests exactly one message. The scenario, 
where each receiver requests more than one message, is discussed in Section~\ref{subsec:multiplerequest}. 
 
Let $\X_0 \subseteq [n]$ and $\vu \in \fqn$. 
In the sequel, we write $\vu \lhd \X_0$ if for any $u_i \neq 0$ it holds $i \in \X_0$. 
Intuitively, this means that if some receiver knows $x_i$ for all $i \in \X_0$ (and also knows $\vu$), then 
this receiver is also able to compute the value of $\vu \cdot \vx$.  

In this paper we consider \emph{linear} index coding. In particular, we assume 
that $S$ broadcasts a vector of $k \in \bbN$ linear combinations 
$\vs = (s_1, s_2, \ldots, s_k) \in \fq^k$, each combination is of the form
\[
s_j = c^{(j)}_1 x_1 + c^{(j)}_2 x_2 + \cdots + c^{(j)}_n x_n = \vcj \cdot \vx \; , 
\]
for $j \in [k]$, where $\{ \vcj = (c^{(j)}_1, c^{(j)}_2, \ldots, c^{(j)}_n)\}_{j \in [k]}$
is a linearly independent set of vectors in $\fqn$.  
Let the code $\C$ of length $n$ and dimension $k$ over $\fq$ be defined as 
\[
\C \define \text{span}(\{\vcone,\vctwo,\ldots,\vck\}) \; . 
\] 
Hereafter, we assume that the sets $\X_j$ for $j \in [m]$ are known to $S$.
Moreover, we also assume that the code $\C$ is known to each receiver $R_j$, $j \in [m]$. In practice 
this can be achieved by a preliminary communication session, when the knowledge of the sets 
$\X_j$ for $j \in [m]$ and of the code $\C$ are disseminated between the participants of 
the scheme. 

The following lemma was formulated in \cite{Yossef} for the case where $\fq$ is a binary field. This lemma specifies a sufficient condition on $\C$ so that the coding scheme is successful, i.e. any $R_j$ has enough data to reconstruct $x_{f(j)}$, 
$j \in [m]$, at the end of the communication session. 
We reproduce this lemma (for the general $\fq$) with its proof for the sake of completeness of the presentation.  
\medskip
\begin{lemma}
\label{lem1}
Let $\mc$ be an $[n,k]_q$-code and let $\{\vcone,\vctwo,\ldots,\vck\}$ be a basis of $\mc$. 
Suppose $S$ broadcasts vector $\vs = (s_1, s_2, \ldots, s_k) = 
(\vcone \cdot \vx, \vctwo \cdot \vx, \ldots, \vck \cdot \vx)$. 
Then, for each $j \in [m]$, the receiver $R_j$ can reconstruct $x_{f(j)}$ if 
the following two conditions hold: 
\begin{enumerate}
\item
there exists $\vu \in \fqn$ such that $\vu \lhd \X_j$; 
\item 
the vector $\vu + \ve_{f(j)}$ is in $\mc$. 
\end{enumerate}
\end{lemma}
\medskip

\begin{proof}
Assume that $\vu \lhd \X_j$ and $\vu + \ve_{f(j)} \in \mc$. 
Since $\vu + \ve_{f(j)} \in \mc$, we obtain that
there exist $\be_1, \be_2, \ldots, \be_k \in \fq$ such that 
\[
(\vu + \ve_{f(j)}) + \sum_{j = 1}^k \be_j \vcj = \vO \; . 
\]
By multiplying by $\vx$, we obtain that 
\begin{eqnarray}
&& \hspace{-8ex} (\vu + \ve_{f(j)}) \cdot \vx + \sum_{j = 1}^k \be_j \big( \vcj \cdot \vx \big) \nonumber \\
& = & (\vu + \ve_{f(j)}) \cdot \vx + \sum_{j = 1}^k \be_j s_j \; = \; \vO \; . 
\end{eqnarray}
Therefore, 
\[
x_{f(j)} = - \sum_{j = 1}^k \be_j s_j - \vu \cdot \vx \; .
\]
Observe that $R_j$ is able to find $\vu$ and all $\be_j$ from the knowledge of the code $\mc$.
Moreover, $R_j$ is also able to compute $\vu \cdot \vx$ since $\vu \lhd \X_j$.
Therefore, $R_j$ is able to compute $x_{f(j)}$. 
\end{proof} 
\medskip

Lemma~\ref{lem1} suggests that in order for the receivers to recover their desired symbols, 
$S$ can use the code $\mc = \text{span}( \{ \vvj + \ve_{f(j)} \}_{j \in [m]})$, for some $\vvj \lhd \X_j$, $j \in [m]$. We show later in Corollary \ref{coro1} that $S$ \emph{must} use a code of such form to guarantee a successful communication session. Finding the lowest dimension code by careful selection of $\vvj$'s is a difficult task (in fact it is NP-hard to do so, see \cite{Yossef, Peeters96}), which, however, yields a scheme with the minimal number of transmissions. 

\subsection{Receivers with Multiple Requests} 
\label{subsec:multiplerequest}
Consider a more general ICSI problem where each receiver requests more than one message. This problem was discussed in~\cite{BirkKol98}. It was shown therein that there exists an equivalent problem with one requested message per each receiver. 
This new problem is easily obtained by splitting each receiver, which requests $\rho>1$ messages, 
into $\rho$ different receivers with the same side information, where each receiver requests exactly one message.
For more detail, the reader can refer to~\cite{BirkKol98}.
In the sequel, we consider scenarios, 
where each receiver requests exactly one message.

\subsection{Scalar and Vector Solutions}

The type of linear solutions considered in this model are referred to as \emph{scalar linear solutions} 
in~\cite{Rouayheb2008, Alon}. For \emph{vector linear solutions}, each message is divided into several packets, 
each packet is a symbol in $\fq$, and a coding scheme combines packets from different messages 
to minimize the number of transmissions. 
It was shown in~\cite{Alon} (see also~\cite{Rouayheb2009},~\cite{Rouayheb2008}) that there 
exist instances of the problem in which a vector linear solution has significantly higher transmission rate 
than any scalar linear solution. Here, the transmission rate of a scheme is defined as the number of packet transmissions 
required for delivery of one packet to each receiver. 

However, if each message consists of $\rho$ packets (symbols in $\fq$), 
then a vector linear solution of this instance can be regarded as 
a scalar linear solution (over $\fq$) of another instance of the index coding problem, 
where each receiver requests exactly $\rho$ messages in $\fq$. 
This instance, in turn, is equivalent to an instance of the ICSI problem considered in this paper. 
Therefore, the two identical $q$-ary linear codes of length $n$ can be used for 
these two equivalent ICSI problems. In that case, in order to study the security of the vector linear solution
of an instance of the ICSI problem, it is enough 
to study the security of the equivalent scalar linear solution of some other instance of the ICSI problem.

\section{Block Secure Linear Index Coding}
\label{sec:our-results}
\subsection{Block Security and Weak Security}

In Section~\ref{sec:our-results}, we present our main results. 
Hereafter, we assume the presence of an adversary $A$ who can listen to all transmissions. 
Let $\mc$ be an $[n,k]_q$-code, and $\{\vcone,\vctwo,\ldots,\vck\}$ be a basis of $\mc$. Let $\vG$ be a generator matrix of $\mc$ whose rows are $\vcone,\vctwo,\ldots,\vck$. Suppose $S$ broadcasts $\vs = (s_1,s_2,\ldots,s_k)=(\vcone \cdot \vx, \vctwo \cdot \vx,\ldots,\vck \cdot \vx)$. The adversary is assumed to possess side information $\{x_i \}_{i \in \X_A}$, 
where $\X_A \subset [n]$. 
For short, we say that $A$ knows $\vx_{\X_A}$. 
The strength of an adversary is defined to be $|\X_A|$. 
Denote $\widehat{\X}_A \define \left( [n] \backslash \X_A \right) \neq \varnothing$. 
Note that from listening to $S$, the adversary also knows $\vs^T = \vG \vx^T$. We define below several levels 
of security for ICSI schemes. 

\medskip
\begin{definition}
\label{defSecurity}
Consider an ICSI scheme, which is based on a linear code $\mc$. The sender $S$ possesses a vector of messages $\vx \in \fq^n$, which is a realized value of a random vector $\vX$.  An adversary $A$ possesses $\{x_i\}_{i \in \X_A}$. 
\begin{enumerate}
\item 
For $B \subseteq \widehat{\X}_A$, the adversary is said to \emph{have no information about $\vx_B$} if 
\begin{equation}
\label{defnoinfo}
\entropy_2(\vX_B|\vG \vX^T,\vX_{\X_A}) = \entropy_2(\vX_B) \; . 
\end{equation}
In other words, despite the partial knowledge on $\vx$ that the adversary has (his side information and the symbols he overheard), the symbols $\vx_B$ still looks completely random to him.  
\item The scheme is said to be \emph{$b$-block secure against $\X_A$} if for every $b$-subset $B \subseteq \widehat{\X}_A$, the adversary has no information about $\vx_B$.  
\item 
The scheme is said to be \emph{$b$-block secure against all adversaries of strength $t$} ($0 \le t \le n-1$) if it is $b$-block secure against $\X_A$ for every $\X_A \subset [n]$, $|\X_A| = t$. 
\item  
The scheme is said to be \emph{weakly secure against $\X_A$} if it is $1$-block secure against $\X_A$. In other words, after listening to all transmissions, the adversary has no information about each particular message that he does not possess in the first place. 
\item
The scheme is said to be \emph{weakly secure against all adversaries of strength $t$ } ($0 \le t \le n-1$) if it is weakly secure against $\X_A$ for every $t$-subset $\X_A$ of $[n]$. 
\item
The scheme is said to be \emph{completely insecure against $\X_A$} if an adversary, who possesses $\{x_i \}_{i \in \X_A}$, by listening to all transmissions, is able to determine $x_i$ for all $i \in \widehat{\X}_A$.
\item 
The scheme is said to be \emph{completely insecure against any adversary of strength $t$} ($0 \le t \le n-1$) if an adversary, who possesses an arbitrary set of $t$ messages, is always able to reconstruct all of the other $n-t$ messages after listening to all transmissions.
\end{enumerate}
\end{definition}
\medskip 

Even when the scheme is $b$-block secure ($b \geq 1$) as defined above, the adversary is still able to obtain information about dependencies between various $x_i$'s in $\widehat{\X}_A$ (but he gains no information about any group of $b$ particular messages). This definition of $b$-block security is a generalization of that of weak security (see~\cite{Bhattad},~\cite{Silva}). Obviously, if a scheme is $b$-block secure against $\X_A$ ($b\geq 1$) then it is also weakly secure against $\X_A$, but the converse is not always true. 

\subsection{Necessary and Sufficient Conditions for Block Security} 

In the sequel, we consider the sets $B \subseteq [n]$, $B \neq \varnothing$, and $E \subseteq [n]$, $E \neq \varnothing$. Moreover, we assume that the sets $\X_A$, $B$, and $E$ are disjoint, and that they form a partition of $[n]$, namely $\X_A \cup B \cup E = [n]$. In particular, $\widehat{\X}_A = B \cup E$. 

\medskip
\begin{lemma}
\label{lem2}
Assume that for all $\vu \lhd \X_A$ and for all $\al_i \in \fq$, $i \in B$ (not all $\al_i$'s are zeros), 
\begin{equation}
\label{E1}
\quad \vu + \sum_{i \in B}\al_i\ve_{i} \notin \mc \; . 
\end{equation}
Then,
\begin{enumerate} 
\item for all $i \in B$: 
\begin{equation}
\vG_i \in \spn(\{\vG_j\}_{j \in E}) \; ; 
\label{eq:span}
\end{equation}
\item 
the system
\begin{equation}
\label{E2}
\vG_E \vy^T = \vG_B \bw^T 
\end{equation}
has at least one solution $\vy \in \fq^{|E|}$ for every choice of $\bw \in \fq^{|B|}$. 
\end{enumerate}
\end{lemma}
\medskip
\begin{proof}
\begin{enumerate}
\item
If $\rank(\vG_E) = k$ then the first claim follows immediately. Otherwise, assume that $\rank(\vG_E) < k$. 
As the $k$ rows of $\vG_E$ are linearly dependent, there exists $\vy \in \fq^k\backslash \{0\}$ such that $\vy \vG_E = 0$. 
\begin{itemize}
\item
If for all such $\vy$ and for all $i \in B$ we have $\vy \vG_i = 0$, then $\vG_i \in ((\spn(\{\vG_j\}_{j \in E}))^\perp)^\perp = \spn(\{\vG_j\}_{j \in E})$ for all $i \in B$. 
\item
Otherwise, there exist $\vy \in \fq^k$ and $i\in B$ such that $\vy \vG_E = 0$ and $\vy \vG_i \neq 0$. Without loss of generality, assume that $\vG = (\vG_{\X_A}|\vG_B|\vG_E)$. Let $\vc = \vy \vG \in \mc$. Then
\[
\vc = (\vc_{\X_A}|\vc_B|\vc_E) = (\vy \vG_{\X_A}| \vy \vG_B | \vy \vG_E ) \; . 
\]
Hence $\vc_B = \vy \vG_B \neq 0$ and $\vc_E = \vy \vG_E = \vO$. Let $\vu = (\vc_{\X_A}|\vO|\vO) \lhd \X_A$ and $\al_i = c_i$ for all $i \in B$. Then $\al_i$'s are not all zero and $\vu + \sum_{i \in B}\al_i \ve_i = \vc \in \mc$, which contradicts (\ref{E1}). 
\end{itemize}
\item
By~(\ref{eq:span}), each column of $\vG_B$ is a linear combination of columns of $\vG_E$. Hence $\vG_B \bw^T$ is also a linear combination of columns of $\vG_E$. Therefore, (\ref{E2}) has at least one solution. 
\end{enumerate}
\end{proof}
\medskip

The following lemma provides us with a criteria to decide whether a particular scheme (based on a code $\mc$) 
is block secure against the adversary $A$ or not. 
This lemma is a generalization of Lemma~\ref{lem1} in the following senses. 
First, while Lemma~\ref{lem1} does not discuss security, observe that the adversary can be viewed as one of the receivers. 
Then, the sufficient conditions (1) and (2) in Lemma~\ref{lem1} (when applied to $A$ and $\X_A$) 
are also sufficient conditions for successful reconstruction 
of a symbol by $A$. Below, we show that these conditions are also necessary. 
Additionally, in the lemma below, 
the weak security, implied by the aforementioned generalization, is further extended to block security. 
(Note that similar statement can be formulated with respect to receivers $R_j$.)

\medskip
\begin{lemma}
\label{lem3} 
For a subset $B \subseteq \widehat{\X}_A$, the adversary, after listening to all transmissions, has no information about $\vx_B$ if and only if
\begin{equation}
\label{E3}
\begin{split}
\forall \vu \lhd \X_A,\ &\forall \al_i \in \fq \text{ with } \al_i,i\in B, \text{ not all zero}:\\ &\vu + \sum_{i \in B}\al_i\ve_{i} \notin \mc \; . 
\end{split}
\end{equation}
In particular, for each $i \notin \X_A$, $A$ has no information about $x_i$ if and only if 
\[
\forall \vu \lhd \X_A \; : \vu + \ve_i \notin \mc \;  . 
\]
\end{lemma}
\medskip

\begin{proof}
Assume that~(\ref{E3}) holds. 
We need to show that the entropy of $\vX_B$ is not changed given the knowledge of $\vG \vX^T$ and $\vX_{\X_A}$. Hence, as shown in Section \ref{sec:preliminaries}, it suffices to show that for all $\bg \in \fq^{|B|}$:
\begin{equation}
\label{E4}
\text{Pr}(\vX_B = \bg| \vG \vX^T = \vs^T,\ \vX_{\X_A} = \vx_{\X_A})= \dfrac{1}{q^{|B|}} \; . 
\end{equation}
Consider the following linear system with the unknown $\vz \in \fq^n$
\begin{equation*}
\begin{cases} \vz_B = \bg \\ \vz_{\X_A} = \vx_{\X_A} \\ \vG \vz^T = \vs^T \end{cases}  , 
\end{equation*}
which is equivalent to 
\begin{equation}
\label{E5}
\begin{cases} \vz_B = \bg\\ \vz_{\X_A} = \vx_{\X_A} \\ \vG_E \vz_E^T = \vs^T {-} \vG_B \bg^T {-} \vG_{\X_A} \vx_{\X_A}^T  \end{cases} \; . 
\end{equation}

In order to prove that (\ref{E4}) holds, it suffices to show that for all choices of $\bg \in \fq^{|B|}$, (\ref{E5}) always has the same number of solutions $\vz$. Notice that the number of solutions $\vz$ of (\ref{E5}) is equal to the number of solutions $\vz_E$ of 
\begin{equation}
\label{E6}
\vG_E \vz_E^T = \vs^T - \vG_B \bg^T - \vG_{\X_A} \vx_{\X_A}^T \; , 
\end{equation}
where $\vs$, $\bg$, and $\vx_{\X_A}$ are known. For any $\bg \in \fq^{|B|}$, if (\ref{E6}) has a solution, then it has exactly $q^{|E|-\rank(G_E)}$ different solutions. Therefore, it suffices to prove that (\ref{E6}) has at least one solution for every $\bg \in \fq^{|B|}$. 

Since $\vx$ is an obvious solution of (\ref{E5}), we have
\begin{equation}
\label{E7}
\vG_E \vx_E^T = \vs^T - \vG_B \vx_B^T - \vG_{\X_A} \vx_{\X_A}^T \; . 
\end{equation}
Subtract (\ref{E7}) from (\ref{E6}) we obtain
\[
\vG_E(\vz_E^T - \vx_E^T) = \vG_B(\vx_B^T - \bg^T) \; ,
\]
which can be rewritten as
\begin{equation}
\label{E8}
\vG_E \vy^T = \vG_B \bw^T \;  ,
\end{equation}
where $\vy \define \vz_E - \vx_E$, $\bw \define \vx_B - \bg$. Due to Lemma \ref{lem2}, (\ref{E8}) always has a solution $\vy$, for every choice of $\bw$. Therefore (\ref{E6}) has at least one solution for every $\bg \in \fq^{|B|}$. 

Now we prove the converse. Assume that (\ref{E3}) does not hold. Then there exists $\vu \lhd \X_A$ and $\al_i \in \fq$, $i \in B$, where $\al_i$'s, $i \in B$ are not all zero, such that
\[
\sum_{i \in B} \al_i \ve_i = \vc - \vu \; ,
\] 
for some $\vc \in \mc$. Hence, similar to the proof of Lemma \ref{lem1}, the adversary obtains
\begin{eqnarray*}
\sum_{i \in B} \al_i x_i & = & \left( \sum_{i \in B} \al_i \ve_i \right) \cdot \vx \\
& = & (\vc - \vu) \cdot \vx \\
& = & \vc \cdot \vx - \vu \cdot \vx \; . 
\end{eqnarray*}
Note that the adversary can calculate $\vc \cdot \vx$ from $\vs$, and can also find $\vu \cdot \vx$ based on his own side information. Therefore, $A$ is able to compute a nontrivial linear combination of $x_i$'s, $i \in B$. Hence the entropy $\entropy_2(\vX_B|\vG \vX^T, \vX_{\X_A}) < \entropy_2(\vX_B)$. Thus, the adversary has some information about the $\vx_B$. 
\end{proof}
\medskip

We have the following straight-forward corollary. 
It generalizes Lemma~\ref{lem1} by providing both the necessary and sufficient conditions 
for the weak security. (Note, that this corollary considers the receiver $R_j$ rather than the adversary $A$, 
since the arguments of Lemma~\ref{lem3} also apply to all receivers.)

\medskip
\begin{corollary}
\label{coro1}
For each $j \in [m]$, the receiver $R_j$ can reconstruct $x_{f(j)}$, $f(j) \notin \X_j$, if and only if
\begin{enumerate}
\item
there exists $\vu \in \fqn$ such that $\vu \lhd \X_j$; 
\item 
the vector $\vu + \ve_{f(j)}$ is in $\mc$.
\end{enumerate}
\end{corollary}
\medskip

Corollary \ref{coro1} suggests that in order for the receivers to recover their desired messages, it is necessary and sufficient to employ a code $\mc$ of the form $\mc = \text{span}( \{ \vvj + \ve_{f(j)} \}_{j \in [m]})$, for some $\vvj \lhd \X_j$, $j \in [m]$. 
\medskip

\begin{theorem}
\label{mainthm1}
Suppose that the source $S$ broadcasts 
\[
\vs = (\vcone \cdot \vx, \vctwo \cdot \vx, \ldots, \vck \cdot \vx) \; , 
\]
where $\{\vcone, \vctwo, \ldots, \vck\}$ is a basis of 
$\mc = \text{span}(\{\vvj + \ve_{f(j)}\}_{j \in [m]})$, 
for some $\vvj \lhd \X_j$, $j \in [m]$. 
Let $d$ be the minimum distance of $\mc$. Then 
\begin{enumerate}
\item The scheme is $(d-1-t)$-block secure against all adversaries of strength $t \le d-2$.
In particular, the scheme is weakly secure against all adversaries of strength $t = d-2$.
\item The scheme is not weakly secure against at least one adversary of strength $t = d-1$. More generally, if there exists a codeword of weight $w$, then the scheme is not weakly secure against at least one adversary of strength $t = w - 1$. 
\item Every adversary of strength $t \leq d - 1$ is able to determine a list of $q^{n-t-k}$ vectors in $\fq^n$ which includes the vector of messages $\vx$.   
\end{enumerate}
\end{theorem}
\medskip 

\begin{proof}
\begin{enumerate}
\item
Observe that by Corollary~\ref{coro1}, every $R_j$, $j \in [m]$, can reconstruct $x_{f(j)}$, $f(j) \notin \X_j$.
Assume that $t \leq d - 2$. By Lemma \ref{lem3}, it suffices to show that for every $t$-subset $\X_A$ of $[n]$ and for every $(d-1-t)$-subset $B$ of $\widehat{\X}_A$,  
\begin{equation}
\label{E9}
\begin{split}
\forall \vu \lhd \X_A,\ &\forall \al_i \in \fq \text{ with } \al_i,i\in B, \text{ not all zero}:\\ &\vu + \sum_{i \in B}\al_i\ve_{i} \notin \mc \; . 
\end{split}
\end{equation}
For such $\vu$ and $\al_i$'s, we have $\weight(\vu + \sum_{i \in B}\al_i\ve_{i}) \leq \weight(\vu) + \weight(\sum_{i \in B}\al_i\ve_{i}) \leq t + (d-1-t) = d -1 < d$. Moreover, as $\supp(\vu) \cap B = \varnothing$ and $\al_i$'s, $i \in B$, are not all zero, we deduce that $\vu + \sum_{i \in B}\al_i\ve_{i} \neq 0$. We conclude that $\vu + \sum_{i \in B}\al_i\ve_{i} \notin \mc$.

\item
We now show that the scheme is not weakly secure against at least one adversary of strength $t = d-1$. The more general statement can be proved in an analogous way. 

Pick a codeword 
$\vc = (c_1, c_2, \ldots, c_n) \in \mc$ such that $\weight(\vc)=d$ and let $\supp(\vc) = \{i_1,i_2,\ldots,i_d\}$. Take 
$\X_A = \{i_1,i_2,\ldots,i_{d-1}\}$, $|\X_A| = d-1$. Let 
$$
\vu = (\vc / c_{i_d} - \ve_{i_d}) \; .
$$
Then, $\vu \lhd \X_A$ and $\vu + \ve_{i_d} = \vc / c_{i_d} \in \mc$. By Lemma \ref{lem1}, $A$ is able to determine $x_{i_d}$. Hence  
the scheme is not weakly secure against the adversary $A$, who knows $d-1$ messages $x_i$'s in advance.

\item
Consider the following linear system of equations with unknown $\vz \in \fq^n$
\begin{equation*}
\hspace{-15ex} \begin{cases}
\vz_{\X_A} = \vx_{\X_A} \\ 
\vG \vz^T = \vs^T
\end{cases} 
, 
\end{equation*}
which is equivalent to
\begin{equation}
\begin{cases}
\vz_{\X_A} = \vx_{\X_A}\\ 
\vG_{\widehat{\X}_A} \vz^T_{\widehat{\X}_A} = \vs^T - \vG_{\X_A}\vx_{\X_A}^T
\end{cases}
\label{E10} . 
\end{equation}
The adversary $A$ attempts to solve this system. 
Given that $\vs$ and $\vx_{\X_A}$ are known, the system~(\ref{E10}) has $n - t$ unknowns and $k$ equations. Note that $t \leq d -1$, and thus by Theorem~\ref{singleton} we have $n - t \geq n - d + 1 \geq k$. If $\rank(\vG_{\widehat{\X}_A}) = k$ then (\ref{E10}) has exactly $q^{n-t-k}$ solutions, as required. 

Next, we show that $\rank(\vG_{\widehat{\X}_A}) = k$. 
Assume, by contrary, that the $k$ rows of $\vG_{\widehat{\X}_A}$, 
denoted by $\mathbf{r}_1,\mathbf{r}_2,\ldots,\mathbf{r}_k$, are linearly dependent. 
Then there exist $\be_i \in \fq$, $i \in [k]$, not all zero, such that $\sum_{i=1}^k \be_i \mathbf{r}_i = 0$.
Let \[
\vc = \sum_{i=1}^k \be_i \cdot \vG[i] \in \mc \backslash \{ \vO \} \; . 
\]
(Recall that $\vG[i]$ denotes the $i$-th row of $\G$). Then $\vc_{\widehat{\X}_A}=\sum_{i=1}^k \be_i \mathbf{r}_i = 0$ and hence $\weight(\vc) = \weight(\vc_{\X_A}) \le t \leq d - 1$. This is a contradiction, which follows from the assumption that 
the $k$ rows of  $\vG_{\widehat{\X}_A}$ are linearly dependent.  
\end{enumerate}
\end{proof}
\medskip

\begin{example}
Let $q = 2$. Assume that $\X_A = \varnothing$ and that $\X_j \ne \varnothing$ for all $j \in [m]$. 
Consider a linear scheme for the ICSI problem, employing an $[n,k,d]_2$-code $\mc$ with $d =2$,
which is defined as follows. For each $j \in [m]$ choose some $i_j \in \X_j$. Let $\mc = \text{span}(\{\ve_{i_j} + \ve_{f(j)}\}_{j \in [m]})$. Then, indeed, $\dist(\mc) =2$. Since $t = |\X_A|=0$, we have $d-1-t = 1$. Therefore by 
Theorem~\ref{mainthm1} the scheme employing $\mc$ is weakly secure against $A$. 
Moreover, if $\mc$ is nontrivial (and so $k < n - d+1$), 
we have $k \leq n - d = n - 2$.
\end{example}

\subsection{Block Security and Complete Insecurity}

Theorem~\ref{mainthm1} provides a threshold for the security level of a scheme that uses a given linear code $\mc$. 
If $A$ has a prior knowledge of any $t \le d - 2$ messages, then the scheme is still secure, i.e. the adversary has no information about any $d-1-t$ particular messages from $\{ x_i \}_{i \in \widehat{\X}_A}$. 
On the other hand, the scheme may no longer be secure against an adversary of strength $t = d-1$. The last assertion of Theorem~\ref{mainthm1} shows us the difference between being block secure and being (strongly) \emph{secure} in a commonly used sense (see, for instance~\cite{CaiYeung2002}). More specifically, if the scheme is (strongly) secure, the messages look completely random to the adversary, i.e. the probability to guess the correct messages is $1/q^n$. However, if the scheme is $(d-1-t)$-block secure (for $t \leq d-2$), then the adversary is able to guess the correct messages with probability $1/q^{n - t - k}$.  

For an adversary of strength $t \ge d$, the security of the scheme depends on the properties of the code employed, in particular, it depends on the weight distribution of $\mc$. From Theorem~\ref{mainthm1}, if there exists $\vc \in \mc$ with $\weight(\vc) = w$, then the scheme is not weakly secure against some adversary of strength $t=w - 1$. In general, the scheme might still be ($b$-block or weakly) secure against some adversaries of strength $t$ for $t \geq d$. While we cannot make a general conclusion on the security of the scheme when the adversary's strength is larger than $d-1$, Lemma \ref{lem3} is still a useful tool to evaluate the security in that situation. However, as the next theorem shows, if the size of $\X_A$
is sufficiently large, then $A$ is able to determine all the messages in $\{ x_i \}_{i \in \widehat{\X}_A}$.   

\medskip
\begin{theorem}
\label{mainthm2}
Suppose that the settings of the coding scheme for the ICSI problem are defined as in Theorem~\ref{mainthm1}. 
Then the scheme is completely insecure against any adversary of strength $t \ge n - d^\perp + 1$, where $d^\perp$ denotes the dual distance of $\mc$.  
\end{theorem}
\medskip 

\begin{proof}
Suppose the adversary knows a subset 
$\{x_i\}_{i \in \X_A}$, $\X_A \subseteq [n]$ and $|\X_A| = t \geq n - d^\perp + 1$. By Corollary~\ref{coro1}, it suffices to show that for all $i \notin \X_A$, there exists $\vu \in \fqn$ satisfying simultaneously $\vu \lhd \X_A$ and $\vu + \ve_i \in \mc$. 

Indeed, take any $i \in \widehat{\X}_A$, and let $\rho = n - t \le d^\perp -1$. 
Consider the $\rho$ indices which are not in $\X_A$. By Theorem~\ref{thmOA}, there exists a codeword $\vc \in \mc$ with 
\begin{equation*}
c_\ell = \begin{cases} 1 &\mbox{ if } \ell = i, \\ 0 & \mbox{ if } \ell \notin \X_A \cup \{ i \} 
\end{cases} \; . 
\end{equation*}
Then $\text{supp}(\vc) \subseteq \X_A \cup \{ i \}$.
We define $\vu \in \fqn$ such that $\vu \lhd \X_A$, as follows. For $\ell \in \X_A$, we set $u_\ell = c_\ell$, 
and for $\ell \notin \X_A$, we set $u_\ell = 0$. 
It is immediately clear that $\vc = \vu + \ve_i$. 
Therefore, by Corollary~\ref{coro1}, the adversary can reconstruct $x_i$. 
We have shown that the scheme is completely insecure against an arbitrary set $\X_A$ satisfying $|\X_A| \geq n - d^\perp + 1$, hence completing the proof. 
\end{proof}
\medskip

When $\mc$ is an MDS code, we have $n - d^\perp + 1 = d - 1$, and hence the two bounds established 
in Theorems~\ref{mainthm1} and~\ref{mainthm2} are actually tight. 
The following example further illustrates the results stated in these theorems. 
\medskip
\begin{example}
\label{ex1}
Let $n=7, m =7, q = 2$. Suppose that the receivers have in their possession set of messages as appears in the third column of the table below. Suppose also, that the demands of all receivers are as in the second column of the table. 
 
\begin{equation*}
\begin{array}{|c||c|c|}
\hline
\text{Receiver} & \text{Demand} & \{x_i\}_{i \in \X_j} \\
\hline
\hline
R_1 & x_1 & \{x_6, x_7\} \\
\hline
R_2 & x_2 & \{x_5, x_7\} \\
\hline
R_3 & x_3 & \{x_5, x_6\} \\
\hline
R_4 & x_4 & \{x_5, x_6, x_7 \} \\
\hline
R_5 & x_5 & \{x_1, x_2, x_6\} \\
\hline
R_6 & x_6 & \{x_1, x_3, x_4\} \\
\hline
R_7 & x_7 & \{x_2, x_3, x_6\} \\
\hline
\end{array}
\end{equation*}

For $j =1 ,2, \ldots, 7$, let $\vvj \in \ff_2^7$ such that $\text{supp}(\vvj)=\X_j$.
Assume that this scheme uses the code $\mc = \text{span}(\{\vvj + \ve_j\}_{j \in [7]})$. 
Then the set $\{\vcj \define \vvj + \ve_j\}_{j \in [4] }$ forms a basis for $\mc$. It is easy to see that this $\mc$ is a $[7,4,3]_2$ Hamming code with $d=3$ and $d^\perp = 4$. 

Suppose that $S$ broadcasts the following four bits: 

\begin{center}
$s_1 = (\vvone + \ve_1) \cdot \vx = \vcone \cdot \vx$ , \\
$s_2 =(\vvtwo + \ve_2) \cdot \vx = \vctwo \cdot \vx$  , \\
$s_3 =(\vv^{(3)}+\ve_3) \cdot \vx = \vc^{(3)} \cdot \vx$  , \\
$s_4 =(\vv^{(4)}+\ve_4) \cdot \vx = \vc^{(4)} \cdot \vx$  . \\
\end{center} 

Each $R_j$, $j = 1 ,2 , \ldots, 7$, can compute $(\vvj + \ve_j) \cdot \vx$ by using linear 
combination of $s_1,s_2,s_3,s_4$. Then, each $R_j$ can
subtract $\vvj \cdot \vx$ (his side information) from $(\vvj + \ve_j) \cdot \vx$ to retrieve $x_j = \ve_j \cdot \vx$. 

For example, consider $R_5$. Since 
\[
\left(\vv^{(5)}+\ve_5\right) \cdot \vx = \left( (\vvone+\ve_1)+(\vvtwo+\ve_2) \right) \cdot \vx = s_1 + s_2 \; , 
\]
$R_5$ subtracts $x_1 + x_2 + x_6$ from $s_1 + s_2$ to obtain 
\begin{eqnarray*}
&& \hspace{-6ex} (s_1 + s_2) - (x_1 + x_2 + x_6) \\
& = & (x_1 + x_2 + x_5 + x_6) - (x_1 + x_2 + x_6) \\
& = & x_5 \; . 
\end{eqnarray*}

If an adversary $A$ has a knowledge of a single message $x_i$, then by Theorem~\ref{mainthm1}, $A$ is not able to determine any other message $x_\ell$, for $\ell \neq i$. Indeed, $\dist(\mc) = 3$, while $t = 1$. Therefore, the scheme is weakly secure against all adversaries of strength $t =1$. Similarly, if the adversary knows none of the messages in advance, then the adversary has no information about any group of $2$ messages. On the other hand, the scheme is completely insecure against any adversary of strength $t \ge 4$; in that case $A$ is able to recover the remaining $n-t$ messages.    
\end{example}

\subsection{Role of the Field Size}

The following example demonstrates that the use of codes over larger fields might have a positive impact on the security level. More specifically, in that example, codes over large fields significantly enhance the security, compared with the 
codes over small field.

\medskip
\begin{example}
\label{ex2}
Suppose that the source $S$ has $n$ messages $x_1, x_2, \ldots, x_n$. Assume that there are $m < n$ receivers $R_1, R_2, \ldots, R_m$, and each receiver $R_j$ has the same set of side information, $\X_j = \{m + 1, m+2, \ldots, n \}$. Assume also that each $R_j$ requires $x_j$, for $j \in [m]$. 

We can define the ICSI scheme based on the code $\mc$, as above. The code employed in this scheme has dimension at least $m$, since all the vectors $\vvj+\ve_j$, for some $\vvj \lhd \X_j$, $j \in [m]$, are linearly independent. Therefore, the number of transmission required in this scheme is at least $m$, which is equal to the number of transmissions in the trivial solution (just broadcasting each of $x_1, x_2, \ldots, x_m$). 

If we employ a \emph{binary} code $\mc$, for the large values of $n$ the minimum distance $d$ of $\mc$ is bounded from above
by a sphere-packing bound
\[
   d \le 2 n \cdot ( \entropy^{-1}_2( 1 - m/n) - \epsilon) \; ,
\]
where $\epsilon \rightarrow 0$ as $n \rightarrow \infty$. 
Hence the scheme, which uses a binary code, is secure against any adversary of strength $t \le d - 2$. 
It is insecure against some adversaries of strength $t \ge d - 1$. 

There is a variety of stronger upper bounds on the minimum distance of binary codes, such as 
the Johnson bound, the Elias bound, and the McEliece-Rodemich-Rumsey-Welch bound (see~\cite[Chapter 4.5]{roth} for more
details). These bounds provide even stronger bounds on the security of this ICSI scheme, when the scheme is
based on the binary code. 

By contrast, consider an $q$-ary code $\mc$, for $q \geq n + 1$ (we also assume here that all $x_i$ are in $\fq$). There exists a $q$-ary MDS code $\mc$ of length $n$, dimension $m$, and with the minimum distance equal $n - m + 1$ (for example, Reed-Solomon code). By employing this code, the new scheme is secure against all adversaries of strength $t \le n - m - 1$. 
In order to find an appropriate generator matrix for the Reed-Solomon code for the settings of this example, we start with some generator matrix of Reed-Solomon code, and then apply Gaussian elimination to obtain a new generator matrix of the form $\mathbf{G} = (\mathbf{I}_m | \mathbf{P})$, where $\mathbf{P}$ is a $m \times (n - m)$ matrix over $\fq$. 

It is well known that there is a significant gap between the Singleton bound and the sphere-packing bound 
(see~\cite[p.~111]{roth} for details). 
Therefore, for some ICSI instances, 
coding over large fields can provide significantly higher levels of security than  
binary coding. 
\end{example}

\section{Conclusion and Open Questions}
\label{sec:conclusion}
In this paper, we analyze the levels of security of linear solutions for the ICSI problem and establish two new bounds. These bounds employ the minimum distance and the dual distance of the linear code used in the scheme. While the dimension of 
this code corresponds to the number of transmissions in the scheme, the minimum distance is related to the security 
of the scheme. The generating matrix of the code depends on the sets of messages that each receiver owns. 

However, there are various generating matrices that can be used for the same instance of ICSI problems. Moreover, 
puncturing of some nonzero entries in the generating matrix, could probably lead to a code with a better minimum 
distance, which in turn corresponds to a ICSI scheme with stronger security. 
Thus, the question which remains open is how to design a code for a particular instance of the ICSI problem, which has 
the largest possible minimum distance. It is very likely that finding such a code is a hard problem. For comparison,
even finding the minimum distance of a code given by its generating matrix is known to be NP-hard~\cite{Vardy1997}.

The following simple generalization of the ICSI problem is called Network Coding with Side Information (NCSI) problem. 
Consider a network with a sender $S$, possessing $n$ messages, 
and $m$ receivers $R_1, R_2,\ldots,R_m$. Each $R_j$ requests one message. Suppose 
that each $R_j$ has some side information, namely $R_j$ knows some subset of these $n$ messages. 
There is also an adversary $A$, listening to some links in the network, who possesses some of the messages. 
Given an instance of the NCSI problem, the following questions arise: 
\begin{enumerate}
\item
Is it possible to satisfy all the requests simultaneously by a single transmission, using linear network coding? 
\item
If there exists network coding solution, how secure is it? 
\end{enumerate}
Some techniques, presented in this paper, can be extended to provide sufficient (and sometimes necessary) conditions for an existence of a linear solution for the NCSI problem, and to analyze the level of security of such a solution. We omit the details from this paper. 

\section{Acknowledgements}

The authors would like to thank Fr\'{e}d\'{e}rique Oggier for helpful discussions. This work is 
supported by the National Research Foundation of Singapore (Research Grant
NRF-CRP2-2007-03).

\bibliographystyle{IEEEtran}
\bibliography{BibFile_SecureICSI_NCSI}

\end{document}